\documentclass[useAMS,usenatbib]{aa}  

\usepackage{graphicx}
\usepackage{txfonts}
\usepackage{graphicx} 
\usepackage{times}
\usepackage{epsfig}
\usepackage{epstopdf}
\usepackage{amsfonts}
\usepackage{amsmath}
\usepackage{amsbsy}
\usepackage{bm}
\usepackage{url}
\usepackage{microtype}
\usepackage{rotating}
\usepackage{sidecap}
\usepackage{longtable}
\usepackage{lscape}
\usepackage{rotating}
 \usepackage{array,multirow,graphicx}
\usepackage[export]{adjustbox}
\usepackage{amssymb,xcolor}
\usepackage{colortbl}
\usepackage{booktabs}
\usepackage{url}
\usepackage[utf8]{inputenc}

\begin{document}

   \title{Rethinking the 67 Hz QPO in GRS 1915+105: type-C QPOs at the innermost stable circular orbit}


   \author{S.E. Motta
          \inst{1}
          \and
          T.M. Belloni\inst{1}\fnmsep \thanks{Deceased.}}

   \institute{$^{1}$Istituto Nazionale di Astrofisica, Osservatorio Astronomico di Brera, via E.\,Bianchi 46, 23807 Merate (LC), Italy\\
              \email{sara.motta@inaf.it}
             }

   \date{Received July 1, 2023; accepted January 30, 2024}

 
  \abstract
   {The study of Quasi-Periodic Oscillations (QPO) at low and high frequency in the variability of the high-energy emission from black-hole binaries and their physical interpretation in terms of signatures of General Relativity in the strong-field regime.}
   {To understand the nature of the 67 Hz QPOs observed in the X-ray emission of the peculiar black-hole binary GRS 1915+105 within the general classification of QPOs, and to determine the spin of the black hole in the system by applying the Relativistic Precession Model (RPM).}
   {Within the RPM, the only relativistic frequency that is stable in time over a large range of accretion rates and can be as low as 67 Hz (for a black-hole mass as measured dynamically) is the nodal frequency at the Innermost Stable Circular Orbit (ISCO). In the application of the model, this corresponds to type-C QPOs. Under this assumption, it is possible to measure the spin of the black hole by using the mass of the black hole previously obtained via dynamical measurements. We re-analysed a large number of RossiXTE observations to check whether other timing features confirm this hypothesis.}
   {The identification of the 67 Hz QPO as the nodal frequency at ISCO yields a value of 0.706 $\pm$0.034 for the black hole spin. With this spin, the only two QPO detections at higher frequencies available in the literature are consistent with being orbital frequencies at a radius outside ISCO. The high-frequency bumps often observed at frequencies between 10 and 200 Hz follow the correlation expected for orbital and periastron-precession frequencies at even larger radii.}
   {}


   \maketitle
%

\section{Introduction}

Quasi-periodic oscillations (QPOs) are variability features frequently observed in the X-ray emissions of compact objects undergoing accretion. These oscillations are believed to originate from the innermost regions of the accretion flow. In a power density spectrum (PDS), QPOs manifest as relatively narrow peaks and their centroid frequency can be linked to dynamic motion and/or accretion-related time scales. QPOs have been known for several decades, yet their precise nature remains enigmatic. Numerous models have been proposed to explain their origin, leading to ongoing debates in the field \citep[see][for a recent review]{Ingram2019}.

At low frequencies ($<$ 30 Hz), three ``flavors'' of QPO have been observed in black-hole (BH) X-ray binaries \citep[e.g.,][]{Wijnands1999a,Remillard2002, Casella2004,Casella2005}. The most common of these types of QPO - called Type-C QPOs \citep{Casella2005} - has a centroid frequency that can vary over a broad range ($\sim$0.1-30Hz). Although other models have been proposed \citep[see, e.g,][for a review]{Belloni2014}, type-C QPOs have been often explained as a manifestation of the nodal (or Lense-Thirring\footnote{The Lense-Thirring precession frequency is a weak-field approximation of the nodal precession frequency, see \cite{Stella1998}.}) precession of plasma orbiting around the BH \citep[e.g.,][]{Stella1998, Ingram2009}. They have also been associated to QPOs observed in neutron-star (NS) X-ray binaries, called Horizontal Branch Oscillations (HBO), suggesting that the same process is at work in both classes of systems \citep[][]{Psaltis1999,Casella2005, Motta2017}.
In addition, QPOs with centroid frequencies above hundred Hz (up to $\sim$500~Hz), although very rarely, have been observed from a small number of BH systems \citep[e.g.,][]{Belloni2012}, either isolated or, in an even smaller number of cases, in pairs. 
Such high-frequency oscillations are referred to as high-frequency QPOs (HFQPOs).

Within the Theory of General Relativity (GR), a particle in a bound orbit around a massive object, together with the orbital frequency ($\nu_{\phi}$), has two additional frequencies associated to it, the vertical and radial epicyclic frequencies ($\nu_{\theta}$ and $\nu_{\rm r}$, respectively). From these frequencies, one can easily calculate two additional frequencies associated to the orbit: the nodal precession frequency ($\nu_{\rm nod}$ = $\nu_{\phi}$ - $\nu_{\theta}$), and the periastron precession frequency ($\nu_{\rm per}$ = $\nu_{\phi}$ - $\nu_{\rm r}$). 
The gas accreting onto a BH can be seen as made of orbiting particles and therefore it is possible that the observed characteristic time scales yielded by the QPOs are the products of relativistic effects which can be associated to the aforementioned frequencies. If this association is established, it represents a powerful diagnostic tool for both GR in the strong-field regime, and the physics of accretion. 

Along these lines, the Relativistic Precession Model \citep[RPM][]{Stella1998,Stella1999,Stella1999a} associates the nodal, periastron precession, and orbital frequency with the type-C QPO and the two HFQPOs - lower and upper, respectively - observed in the light curves of accreting BHs and NSs. For BHs, the number of detections of HFQPOs is very low and they are visible together with a type-C QPO even more rarely. 
Despite the scarcity of data, \cite{Motta2014a, Motta2014b, Motta2022} showed that the RPM can be associated to the QPOs and broad noise components observed in the the BH binaries GRO J1655-40, XTE J1550-564 and XTE J1859+226. These works showed that the application of the RPM allows the self-consistent estimate of both spin and mass of a BH from timing features.

The bright BH binary  GRS 1915+105 went into outburst in 1992 and is still active, although in a low-flux state \cite[see, e.g.,][]{Motta2021}. This system is known to be rather peculiar, displaying extreme structured variability on time scales above the second \citep[see][for a review]{Fender2004a}. 
Despite these peculiarities, the PDS from GRS 1915+105 is not unlike that of other BH binaries, and strong type-C QPOs have been observed in its harder states \citep[e.g., ][]{Markwardt1997, Ratti2012}.
At high frequencies,  at variance with other systems, GRS 1915+105 has shown a number of features which have been classified as HFQPOs. In particular, a peak around 67 Hz has been consistently detected with RossiXTE for the sixteen years of operation of the satellite and in the past few years also by Astrosat \citep[see, e.g.,][]{Morgan1997,Belloni2013a, Belloni2019}. This QPO was never observed together with a type-C QPO and represents a very stable frequency in the system.

In this paper we show that the features that until now have been classified as HFQPOs in GRS 1915+105 are instead consistent with being type-C QPOs produced near the innermost stable circular orbit (ISCO). Based on this assumption, we infer the spin of the BH hosted in GRS 1915+105, and we show that the inferred spin, coupled  with the dynamical BH mass measurement, can be used to predict theoretical frequencies that match the data.

\section{Rethinking the 67~Hz QPO}

\cite{Belloni2013a} reported a compilation of all the HFQPOs found in a systematic search performed on all the available RXTE data on GRS 1915+105. The centroid frequencies of 49 of the 51 peaks reported in such a work are found between 58 and 72 Hz (a histogram of such frequencies is shown in Fig. \ref{fig:histoC}). The two remaining peaks, detected at $\sim$134 and 143 Hz,  are discussed in Sec. \ref{sec:HFQPOs}. 

The frequencies of such 49 QPOs are distributed in a narrow range centred around 67~Hz, and henceforth we will refer to these QPOs collectively as \textit{the 67 Hz QPOs}, aware that in reality they span a range of frequencies. Such a narrow range implies that they are determined by a parameter that - along with mass and spin - has to remain unchanged despite large swings in accretion rate. The most obvious candidate is the ISCO, which itself depends only on the mass and spin of the central black hole. The shape of the histogram in Fig. \ref{fig:histoC} support the above fact, as it shows a clear drop at frequencies slightly above 67~Hz.

If we calculate the orbital frequency at ISCO (which at such a radius equals the periastron precession frequency) for a dimensionless spin parameter\footnote{The dimensionless spin parameter is defined as $a^* = J/M^2$, where $J$ is the BH angular momentum and M its mass.} (hereafter spin) a$^*$ = 0  and a mass of 14.4 (i.e. the upper limit on the mass based on the values reported by \citealt{Reid2014}, M = 12.4$^{+2.0}_{-1.8}$M$_{\odot}$) we obtain $\sim$ 152~Hz\footnote{Note that we deliberately ignore solutions where the spin is counter rotating, based on the assumption that the BH spin and the angular momentum of the matter in the accretion disk are unlikely to be anti-parallel \citep[][for a discussion]{Motta2018}.}. This is the minimum possible orbital frequency in GRS 1915+105 at ISCO, as any larger BH spin value and any lower mass will yield higher frequencies. Thus, the 67~Hz QPOs are not consistent with being the result of the orbital motion of matter at the ISCO. 

An intriguing possibility is that the 67~Hz QPO is the result of nodal precession at the ISCO. This would imply that the oscillations observed around 67~Hz are \textit{not} HFQPOs, as it has been believed for decades, but are instead type-C QPOs arising from very close to the ISCO. In this scenario, 67 Hz is the maximum possible centroid frequency for a type-C QPO for a given BH mass and spin. Consequently, the 67 Hz QPOs in GRS 1915+105 represent the high-frequency end of a broad distribution that includes all type-C QPOs in this system, which are most commonly observed below  10 Hz. The fact that a 67 Hz QPO has never been concurrently observed with another type-C QPO supports this hypothesis. This perspective also coherently clarifies why GRS 1915+105, unlike other BH X-ray binaries, exhibits numerous PDS peaks identified as HFQPOs: the 67 Hz QPOs would be reclassified not as HFQPOs, which are relatively rare in BH X-ray binaries, but rather as type-C QPOs, a more frequent occurrence in these systems.

To the best of our knowledge, current literature lacks efforts to interpret the $\sim$67 Hz QPOs in GRS 1915+105 as anything other than HFQPOs \citep{Morgan1997}. Alternative models have been previously proposed \citep[see, e.g.,][]{Fragile2016,Ingram2019} that could offer different interpretations of these QPOs, and which could be tested using the data considered here (all available in the literature). However, conducting such tests falls outside the scope of of this work.

\bigskip 

In the scenario we have outlined, three facts are worth mentioning. Firstly, the 67 Hz QPOs do not exhibit characteristics typical of Type-C QPOs. Secondly, there are additional QPOs identified in GRS 1915+105 with frequencies close yet inconsistent with the 67 Hz QPOs \citep[][and references therein]{Belloni2013}. Finally, at least one other source has been observed to exhibit a QPO at approximately 67 Hz. We discuss each of these points in detail below.

At first glance the 67 Hz QPOs do not show the typical characteristics of type-C QPOs. While type-C QPOs at lower frequencies are generally superimposed on broad-band variability, the 67 Hz QPOs distinctively emerge from the Poisson noise. Additionally, these QPOs exhibit a relatively low rms amplitude (lower than 1\%). These properties should not surprise. Firstly, the broad band variability typical of BH XRBs which always shows a cut-off between a few tenths of Hz and at a few Hz). Consequently, as the frequency of a QPO increases, it becomes increasingly probable for it to arise from Poisson noise rather than from broad-band variability. A similar situation can be observed in GRO J1655-40, where Type-C QPOs observed in the soft state are observed at around 30~Hz in a region of the PDS which is Poisson-noise dominated. Furthermore, the amplitude of type-C QPOs is anti-correlated with the centroid frequency, and thus higher frequency oscillations are expected to show lower amplitudes \citep[see, e.g.,][]{Motta2015}. 

Several authors reported the discovery of various QPO peaks in the PDS from GRS 1915+105, at 27 \citep{Belloni2001}, 34 \citep{Belloni2013} and 41 Hz \citep{Strohmayer2001a}. The 34 Hz QPO can be readily interpreted as harmonically related to the 68 Hz QPO observed in the same PDS. This harmonic relationship is a common feature of Type-C QPOs, which makes the above unsurprising.  Even the occurrence of an isolated 34 Hz QPO (i.e., without a concurrent higher frequency harmonic peak) would not be unexpected. This is due to the variable amplitude of harmonically-related peaks observed in several X-ray binaries, particularly in neutron star systems, where the most prominent, and sometimes the only significant QPO peak does not necessarily correspond to the fundamental frequency of the modulation underlying the signal \citep{Motta2017}. Consequently, the detection of an \textit{orphan} 27 Hz peak might be interpreted in a similar way, presuming an undetected fundamental peak at approximately 54 Hz.
It is more challenging to provide a straightforward explanation for the 41 Hz frequency reported in \cite{Strohmayer2001a}, which was detected concurrently with the 67 Hz QPO, and yet appears to be unrelated to it. We hypothesise that these two QPOs, despite being observed in the same dataset spanning over 15~ks, might not be occurring strictly simultaneously. Under this assumption, the 41 Hz peak  could be a type-C QPO at a lower frequency, with its frequency increasing to 67 Hz at a different time. However, this hypothesis requires further analysis for validation or refutation, a task we reserve for future work.

The 67 Hz QPO is not exclusive to GRS 1915+105 as a similar QPO at 66 Hz has been detected in the candidate BH X-ray binary IGR J17091-3624 during its first observed outburst (2011). Notably, this QPO was observed only once throughout all recorded outbursts of the source, although some excess in the PDS around 66 Hz and 164 Hz was detected in a number of observations of the same outburst \citep{Altamirano2011}. While we cannot exclude this possibility, current evidence does not support the conclusion that this QPO is of the same type as the 67 Hz QPO observed in GRS 1915+105. Nonetheless, it is intriguing that among known BH X-ray transients, the one exhibiting a QPO near 67 Hz shares phenomenological resemblances with those observed in GRS 1915+105 \citep{Altamirano2011, Motta2021}. This similarity warrants further investigation, particularly should any future data confirm that the 66 Hz modulation in IGR J17091-3624 shares properties with the 67 Hz QPOs in GRS 1915+105.

\subsection{The spin}\label{sec:spin}

Under the intriguing hypothesis that the 67 Hz QPOs arise from nodal precession in the proximity of ISCO, it is possible to obtain an estimate of the BH spin as follows. First, we fit the histogram in Fig. \ref{fig:histoC} with a Gaussian function, and we take the Gaussian's peak frequency as the frequency of the QPO arising at ISCO, and the Gaussian's FWHM as the 1-sigma error on such a frequency. This yield $\nu_{nod@ISCO}  =  67.7 \pm  2.1 $ (see Tab. \ref{tab:BHpar}). 

\begin{figure}
\centering
\includegraphics[width=0.43\textwidth]{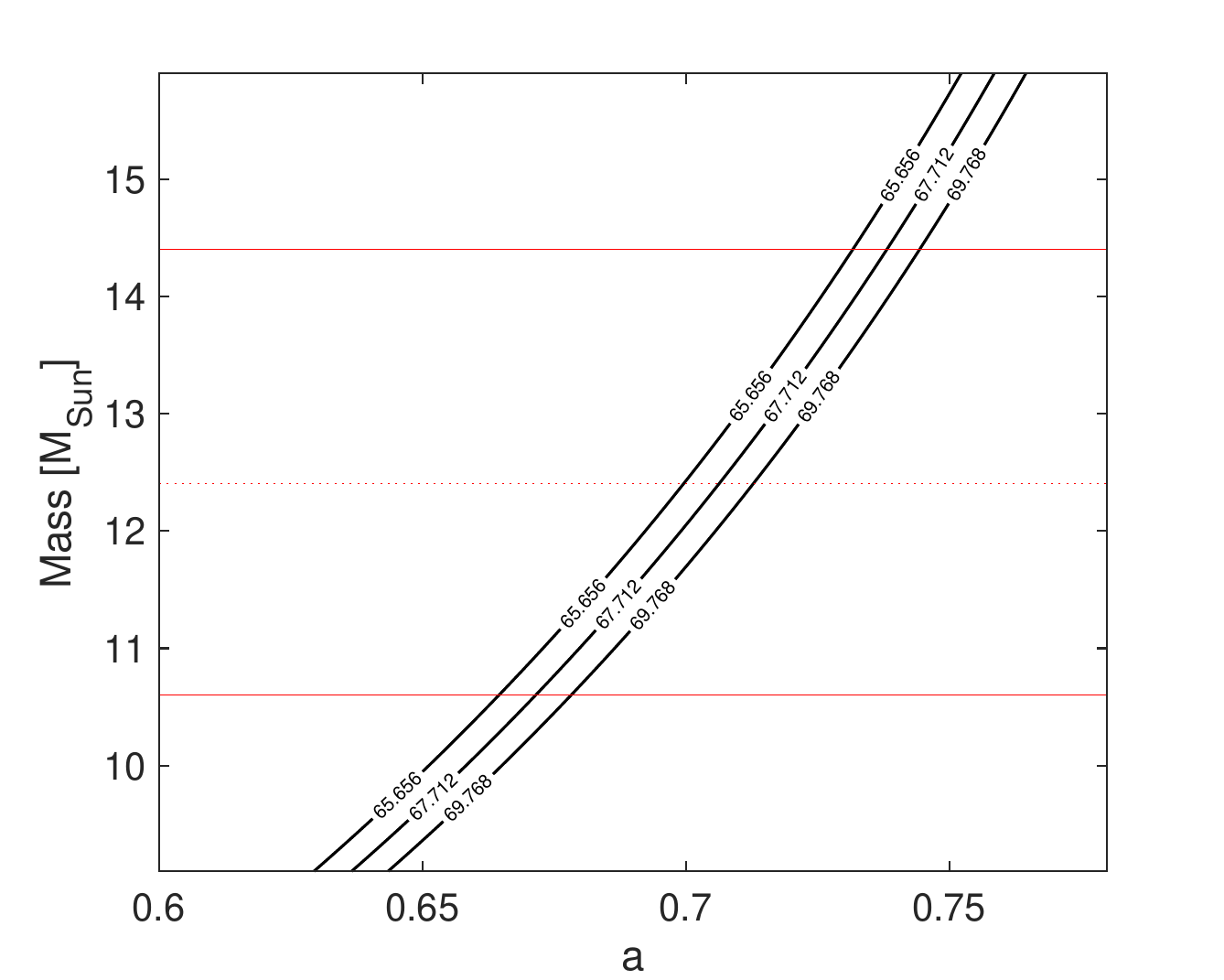}
\caption{Mass versus spin solutions of the RPM obtained assuming that the nodal precession occurs at ISCO. The solutions corresponding to the frequency 67.712 +/- 1-sigma uncertainties are shown. The horizontal solid lines and the dotted line are the upper and lower limit to the dynamical mass, and its central value, respectively  (M = 12.4$^{+2.0}_{-1.8}$M$_{\odot}$). A version of this figure spanning a broader mass and spin range is given in \cite{Franchini2017}.}
\label{fig:mass_spin} 
\end{figure}

Next, we use the RPM to obtain an estimate of the spin. As described in \cite{Franchini2017}, by substituting the expression of the radius of ISCO into the equation of the nodal precession frequency, one can remove the radial dependency and obtain an equations that only depend on the mass and spin of the BH. Adopting a BH mass of 12.4$^{+2.0}_{-1.8}$M$_{\odot}$ we obtain a spin of a$^*$ = 0.706 $\pm$0.034. The corresponding graphical solution is shown in Fig. \ref{fig:mass_spin}. For this spin the radius of the innermost stable circular orbit radius is $R_{\rm ISCO}$ = (3.36$\pm$0.15) $R_{\rm g}$ (see Tab. \ref{tab:BHpar}).

As noted in \cite{Motta2022}, we stress that if the RPM was an exact description of the behaviour of particles orbiting a BH, the uncertainties on the derived on the  BH spin (and mass, and emission radius) would only come from the uncertainties on the measured QPO frequencies, which are dominated by the accuracy of the detecting instrument. However, the exact geometry of the emitting region is unknown, and so is the exact emission mechanism behind QPOs. Both the above facts (and likely others) might be sources of systematics. For instance, in \cite{Motta2022} we gauged the uncertainties that could be related with the the radial extent of the region originating the QPOs, and found that such systematics would be of the order 15\%. Since the effective impact of systematics on the spin estimate is hard to determine accurately, we warn the reader that in the following - for simplicity and to avoid biasses - we will report and use the uncertainties derived uniquely from the errors on the QPO centroid frequencies.

\begin{figure*}
\centering
\includegraphics[width=0.95\textwidth]{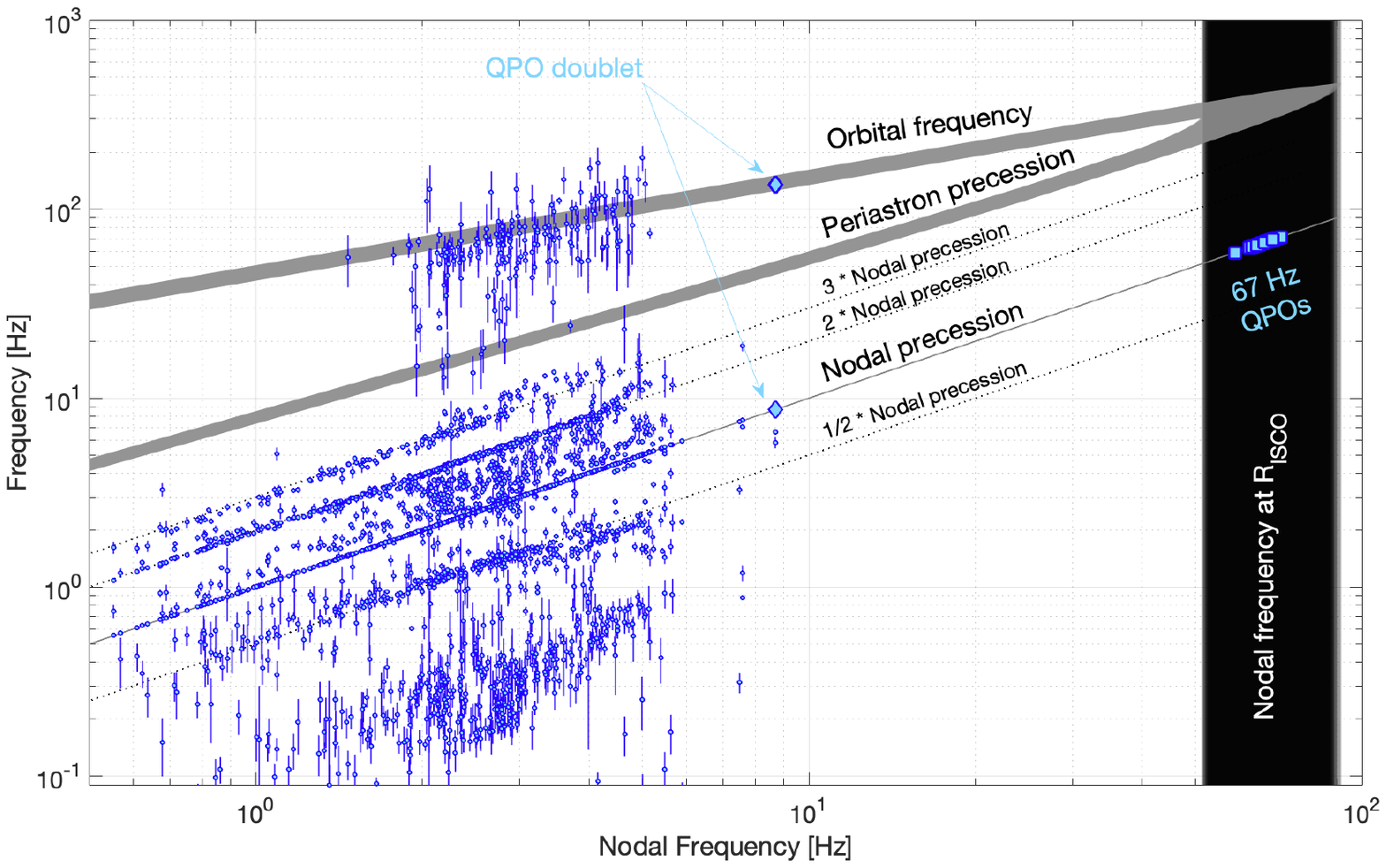}
\caption{Nodal precession frequencies (bottom grey line), periastron precession frequencies (middle lines), and orbital frequencies (top lines) plotted as a function of the nodal precession frequency, as predicted by the RPM using M~=~(12.4$\pm$0.46)~M$\odot$ and a$^*$ = 0.706 $\pm$0.034. The set of lines forming grey regions in the figure each correspond to mass-spin pairs drawn from a bi-variate distribution (see main text ). The lines corresponding to the Nodal precession Frequency overlap to each other as the Nodal precession frequency is the  independent variable in this plot, hence the only difference between lines is in the maximum value they reach (i.e. the nodal precession frequency at ISCO). The dotted lines correspond to the sub-harmonic, the second and third harmonic of the nodal precession frequency. The vertical black lines correspond to the predicted nodal frequency at $R_{\rm ISCO}$ for the same pairs. For clarity we plot only the predicted frequencies for mass-spin pairs within 1-$\sigma$ from the central mass-spin value. The QPO doublet formed by a HFQPO at 135Hz and a type-C QPO at 8.7 Hz are indicated by clear blue diamonds. 
The 67Hz QPOs are marked by clear blue squares. All the PDS broad components described in Section \ref{sec:PBK} are marked by small blue dots with vertical error bars.}
\label{fig:relativistic_frequencies} 
\end{figure*}

\begin{figure}
\centering
\includegraphics[width=0.45\textwidth]{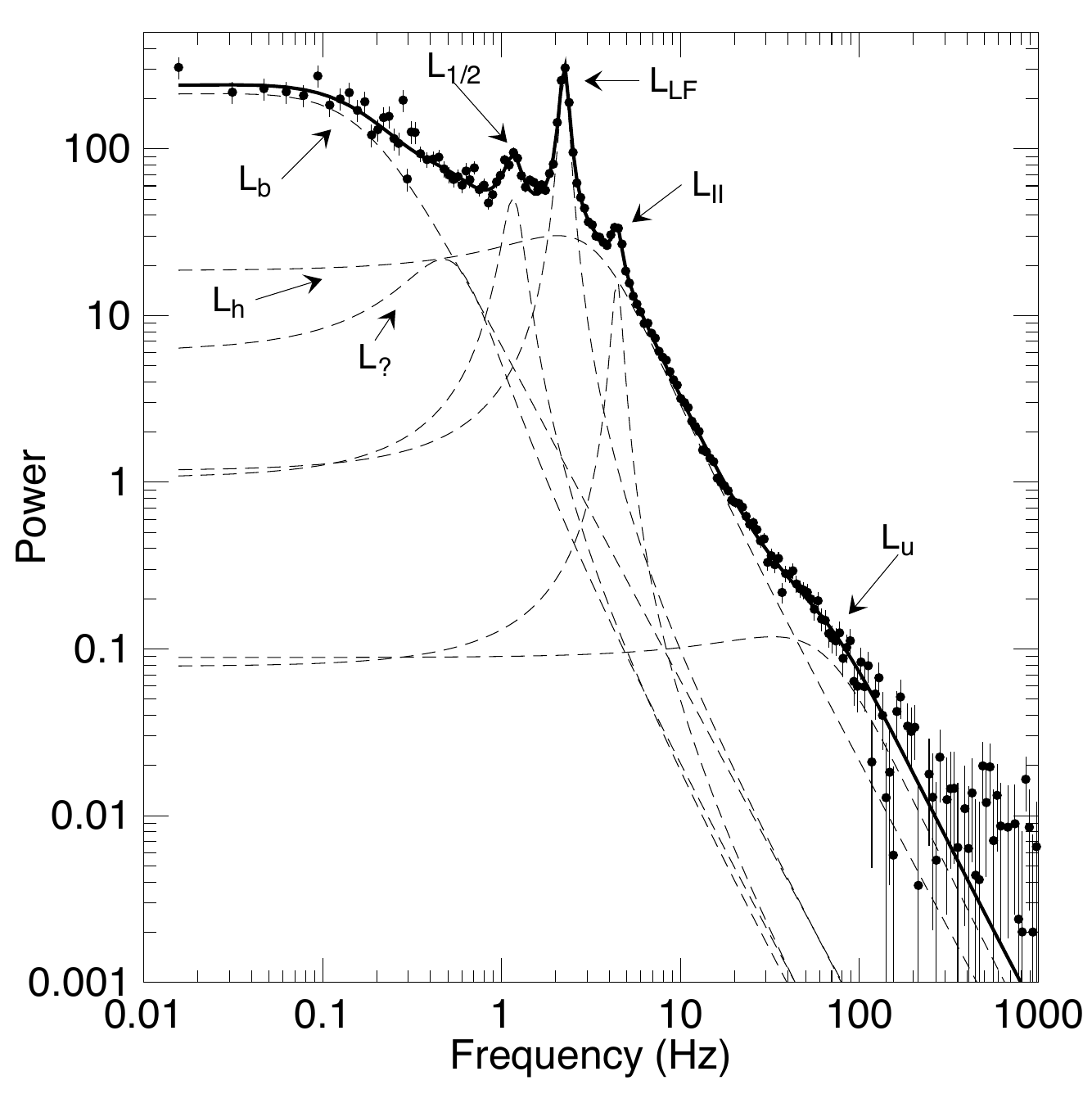}
\caption{Example of a PDS from GRS 1915+105 (RTXE observation 40703-01-01-00). The Lorentzian components fitted to the data are marked with dashed lines. The classifications of the peaks according to \cite{Belloni2002} is shown.}
\label{fig:PDS_components} 
\end{figure}

\section{Completing the picture}
The hypothesis that the 67 Hz QPOs is compatible with being the result of nodal precession near ISCO can be tested by considering the characteristic frequencies of additional components in the PDS, such as the high-frequency broad components often observed with type-C QPOs in BH binaries \citep[][]{Psaltis1999, Stella1999} or, in case they are detected, the frequencies of actual HFQPOs. 

We proceed by predicting the frequencies of the particles orbiting the BH at any radius larger than $R_{\rm ISCO}$ as was done in, e.g., \cite{Motta2014a, Motta2014b, Motta2022}.
In order to take into account the uncertainty associates with the mass and spin of the BH, we simulated a 10$^6$ elements mass and spin normal bi-variate distribution with central value and width equal to the values and uncertainties given in Tab. \ref{tab:BHpar}, and we predict the nodal, periastron precession, and orbital frequencies as a function of radius, as well as the value of $R_{\rm ISCO}$, and the corresponding nodal frequency for mass-spin pairs drawn randomly form the distribution. 

In Figure \ref{fig:relativistic_frequencies} we plot the three set of frequencies as a function of the nodal frequency (grey lines), as well as the values of the nodal frequency at $R_{\rm ISCO}$ calculated in each step, following the procedure adopted in \citealt{Motta2022} (black vertical lines). Every set of grey lines corresponds to a different mass-spin pair drawn from a bi-variate distribution (for clarity we only plot values for the mass-spin pairs within 1-$\sigma$ from the central mass-spin value). The 67Hz QPOs are marked by squares.

\subsection{Bringing in the \textit{real} HFQPOs}\label{sec:HFQPOs}
HFQPOs in BH X-ray binaries are classified solely based on their frequency, which in the majority of systems is of the order 100~Hz. Hence, we classify the QPOs at $\sim$134 and $\sim$143~Hz in GRS 1915+105 reported in \cite{Belloni2013a} as HFQPOs. In this work we will refer to these oscillations as the \textit{real} HFQPOs to distinguish them from the 67~Hz QPOs, which, as already noted, are found at frequencies significantly lower than the typical HFQPO in other BH systems. 

The inspection of the two PDS showing peaks at $\sim$134 and $\sim$143~Hz, respectively, reveals in both cases a clear LFQPO at $\sim$6.5Hz. Such oscillations can be classified as a type-B QPO, based on their frequency and morphology \citep[see, e.g.,][]{Casella2005, Motta2009}. A close inspection of the PDS shows the presence of a weak peak at $\sim$8.54 Hz in one of the two cases, i.e. observation 50703-01-10-02, which according to \cite{Belloni2013a} shows $\rho$-type variability). We classify the weak peak as a type-C QPO because it is not consistent with being the second harmonic to the modulation at 6.5 Hz, and because no type-B QPO has been observed above 7 Hz in GRS 1915+105 \citep[see, e.g., ][]{Soleri2008}. This PDS configuration, i.e. a type-B next to a weak type-C QPO, is relatively uncommon in BH X-ray binaries \citep{Motta2015}, but is typical of neutron star Z-sources \citep[see, e.g.,][]{Motta2019}, which like GRS 1915+105 are believed to accrete near the Eddington limit \citep[see][for the case of Sco X-1]{Motta2017}\footnote{We note that the type-B and type-C QPOs that we detected are found at centroid frequencies in a 4/3 ratio. We believe that the fact that the two peaks are in a 4/3 ratio is pure coincidence as any slightly different rebinning of the PDS results in slightly different centroid frequencies, which are not in a 4/3 ratio.}.

We fitted the PDS as described in several works in the literature, with a number of Lorentzian shapes \citep[e.g.][]{Belloni2002}. The best fit yields a type-C QPO frequency $\nu_{C}$ = 8.71$\pm$ 0.09 and a HFQPO at 135.25$\pm$0.9. A plot showing the fitted PDS is shown in the appendix (Fig. \ref{fig:bc}). \cite{Belloni2013a} already showed that the HFQPO is formally significant after considering the number of trials. The type-C QPO, instead, is borderline significant, i.e. approximately 3.1 $\sigma$ single trial based on the QPO amplitude and its uncertainty.  
Determining the formal significance of a QPO super-imposed to a complex continuum (i.e., not simple red noise) is non trivial \citep[see, e.g.,][]{Vaughan2005}, but for the sake of the argument, we will assume that the type-C QPO is correctly classified and statistically significant, and hence we will treat the tentative type-C QPO plus the HFQPO at $\sim$143~Hz as a QPO ``doublet''. 
Marking this QPO ``doublet'' on the plot in Fig. \ref{fig:relativistic_frequencies} we see that the HFQPO peak is consistent with being a upper HFQPO, and can be associated with the orbital frequency according to the RPM. By extension and in virtue of their similar frequency, we classify also the other high-frequency peak at 143 Hz as a upper HFQPO. 

Following the prescription in \cite{Ingram2014} and by considering the BH mass for GRS 1915+105 from dynamical measurements, we are able to infer an additional, independent value of the spin  to be compared with the value obtained above for consistency. 
We derive a second spin value, a$_{*}^{D}$ = 0.71$\pm$0.13, which is consistent with the value we obtained above in Sec. \ref{sec:spin}. 

\subsection{Bringing in the PDS broad components}\label{sec:PBK}
Finally, we consider the broad PDS components that have been associated with the HFQPOs in BH and NS X-ray binaries thanks to the well-known correlation discovered by \cite{Psaltis1999}, the so-called Psaltis-Belloni-Van der Klis (PBK) correlation. While these broad-band components frequently appear in the PDS of BH X-ray binaries in general and in GRS 1915+105 in particular, their presence is not ubiquitous across all observations. Thus, instead of inspecting all the 1816 RXTE archival observations of GRS 1915+105, we focussed on the sample considered by \cite{Zhang2022}, constructed based on the presence of broad-band signal at high frequencies (or a ``bump''). 

Each observation was reduced following the standard procedures described in several works by our group (one recent example is  \citealt{Motta2022}), with one main difference: since GRS 1915+105 is known for being a variable source, and changes in QPO centroid frequencies can happen on time-scales shorter than the average RXTE observation (a few thousands seconds spread acorss several satellite revolutions), we produced PDS employing 64~s long data segments and generated an average PDS accumulating a fixed number of segments (48) for a total of approximately 3000~s per average PDS, i.e. long enough to provide a good signal-to-noise-ratio in each spectrum, but not so long that the underlying power spectral density distribution varied significantly in the time interval considered, (thus violating the assumption of the Fourier analysis). 

We generated a total of 480 average PDS using custom software under IDL (GHATS\footnote{\url{http://www.brera.inaf.it/utenti/belloni/GHATS_Package/Home.html}}), which we normalised according to \cite{Leahy1983} and we fitted with a combination of Lorentzian components. 
For all the PDS we calculated the characteristic frequency of all the Lorentzians components, defined as $\nu_{\rm max} = \sqrt{\nu^2 + (\Delta/2)^2}$, where $\nu$ and $\Delta$ are the frequency and width, respectively \citep[see][]{Belloni2002}. Note that for narrow features such as QPOs the characteristic frequency is by construction very similar to the Lorentzian peak frequency. 

While the PDS from GRS 1915+105 (and of BH binaries in general) can be complex, and it is often hard to classify components based on a given scheme (i.e., not all the components are all present all the time in every PDS), the majority of the components can be classified as follows, and as it is shown in Fig. \ref{fig:PDS_components}. The PDS can include:
\begin{itemize}
    \item a type-C QPO (L$_{LF}$) and its harmonic content, i.e., a sub-harmonic (L$_{1/2}$), a second harmonic (L$_{II}$), and sometimes a third or fourth harmonic;
    \item a broad Lorentzian at low frequencies (L$_b$);
    \item a second broad Lorentzian peaking at frequency below the PDS break frequency (L$_?$);
    \item a third broad Lorentzian centred at a frequency close to the QPO centroid frequency (L$_h$);
    \item one or two broad Lorentzians at high frequencies (L$_u$) and/or (L$_l$, not shown in the figure).
\end{itemize}  


In the attempt to interpret the PDS components from GRS 1915+105 in an unbiased way, we plotted all the characteristic frequencies we found in all the PDS we generated in Fig. \ref{fig:relativistic_frequencies}, together with the frequencies predicted by the RPM given the dynamical mass from photometry and the spin we calculated in Sec. \ref{sec:spin}. All frequencies increase with the frequency of the type-C QPO (L$_{LF}$).  In particular, in the context of the RPM, the Nodal Precession frequency corresponds to L$_{LF}$, and the sub-harmonic, second and third harmonic to this frequency correspond to L$_{1/2}$, L$_{II}$ and L$_{III}$. The periastron precession frequency and orbital frequency correspond to the broad components detected at high frequencies, L$_{l}$ or L$_{u}$, respectively, or to HFQPOs (not shown in Fig. \ref{fig:PDS_components}), as suggested by \cite{Psaltis1999}.  

In Fig. \ref{fig:PDS_components} several points do not clearly coincide with one of the frequencies of the RPM, and this is because the RPM aims to explain only three components of the several in the PDS. 
What is important to notice here is that several points are consistent with the frequencies predicted by the RPM for the mass and spins we derived from the 67 Hz QPOs under the assumption that they are the result of nodal precession of particles at ISCO. The vast majority of the points that are inconsistent with the RPM frequencies can be associated with the L$_b$, L$_h$ or L$_?$ components (see Fig. \ref{fig:PDS_components}). 

\subsection{Implications}

The spin we obtained for the BH in GRS 1915+105 is the fourth one obtained via X-ray timing. Previous estimates were obtained for the BH X-ray binaries GRO J1655-40, XTE J1550-564, XTE J1859-226 presented in \citealt{Motta2014a}, \citealt{Motta2014b}, \cite{Motta2022}, and the the spins were 0.29, 0.34, and 0.15, respectively. In these cases, the spin values we derived are relatively low if compared to those typically obtained via other `electromagnetic' spin measurement methods (continuum fitting, reflection spectroscopy, and reverberation,  \citealt{Reynolds2021} for a recent review), which so far yielded spin distributions peaked at values above 0.9 (especially true in the case of the reflection-based measurements, see e.g. \citealt{Draghis2023}). Instead - and given the many peculiarities of this system this is perhaps unsurprising - the spin for GRS 1915+105 is relatively high - 0.76 - although still inconsistent with the value from X-ray spectroscopy \citep{Miller2013}. 

While the number of timing-based spin values from the RPM is still low, they  appear to be consistent with the BH spin distribution obtained based on the entire gravitational waves sample GWTC-3 \citep{LIGOVIRGOKAGRA2021}. This distribution peaks around spin $\sim$ 0.2 and shows a thin tail extending to higher values, thus indicating that the vast majority BHs in binary BH systems are relatively slow rotators. 
Unlike the case of other `electromagnetic' methods, the results of which indicate that the spin distributions from binary BHs and BH X-ray binaries may be different \citep{Draghis2023}, our results seem to suggest that the LIGO/Virgo/KAGRA BHs and X-ray binary BHs may feature a similar spin distribution, and hence may be members of strictly related populations \citep[see][]{Belczynski2021, Fishbach2022}.

\begin{table}
\caption{A summary of the quantities measured and derived in this work. Uncertainties are given at a 1-sigma level. }             
\label{tab:BHpar}      
\centering                          
\begin{tabular}{c c c c}        
\hline                        
M                               & = &       12.4$^{+2.0}_{-1.8}$M$_{\odot}$ M$\odot$     & \cite{Reid2014}  \\
\textbf{a$_{*}$}                & = &       \textbf{0.706$\pm$0.034}                     &  From 67Hz QPOs   \\
a$_{*}^{D}$                     & = &       0.71$\pm$0.13                                &  From QPO doublet   \\
$R_{\rm ISCO}$                  & = &       3.36$\pm$0.15 $R_{\rm g}$                    &  Derived    \\
\hline \hline
$\nu_{\rm nod@ISCO}$             & = &       67.712 $\pm$  2.056 Hz                       &  Measured    \\
$\nu_{HFQPO}^{\rm upper}$       & = &       135.250  $\pm$  0.70                         &  Measured    \\
$\nu_{C}$                       & = &       8.72 $\pm$  0.09                             &  Measured    \\
\hline                                   
\end{tabular}
\end{table}

\section{Conclusions}

Building on the observation that the 67 Hz QPOs in GRS 1915+105 are confined to a narrow frequency range, we hypothesized these QPOs to be a consequence of nodal precession at the ISCO around a spinning black hole.

Using the Relativistic Precession Model and the black hole mass determined from dynamical measurements (M = 12.4$^{+2.0}{-1.8}$M${\odot}$), we derived a moderately high black hole spin (a$^*$ = 0.706 $\pm$ 0.034). Our predictions of frequencies around a black hole with this spin and mass nicely matched the observational data, lending further support to the idea that certain PDS features in accreting BH X-ray binaries can be explained by matter motion near a compact object.

We conclude that the 67 Hz QPOs observed in GRS 1915+105, historically classified as HFQPOs, are more plausibly type-C QPOs originating from the vicinity of the ISCO of a moderately spinning black hole.

\begin{acknowledgements}

      The authors acknowledge financial contribution from grant PRIN INAF 2019 n.15.
      This work benefited from discussions during Team Meetings of the International Space Science Institute (Bern), whose support we acknowledge.\\
      SEM heartily thanks Tomaso Belloni, who sadly passed away before this paper could be completed. Among many other contributions to the field, his work was crucial for the understanding of GRS 1915+105.\\
      SEM acknowledge the assistance of ChatGPT for proofreading and language enhancement in the preparation of this manuscript.

\end{acknowledgements}

%
%

\bibliographystyle{aa.bst}
\bibliography{biblio} 

\newpage
\begin{appendix} 

\section{Additional figures}
\begin{figure}
\centering
\includegraphics[width=0.43\textwidth]{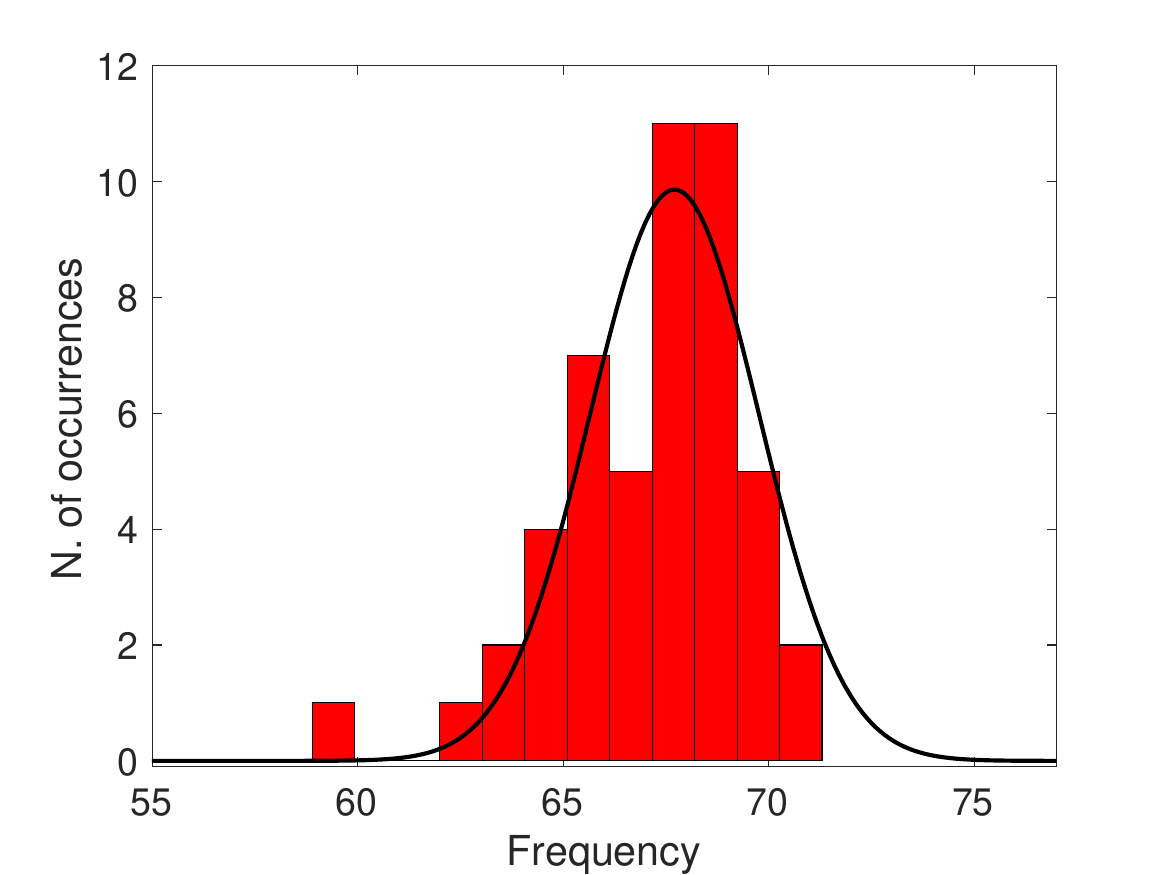}
\caption{A Histograms of the centroid frequencies of the QPOs found around 67~Hz in the data from GRS 1915+105, as reported in \cite{Belloni2013a}. The black solid line is the best Gaussian fit to the histogram, which yields a centroid frequency of 67.712 Hz and a FWHM of 2.056 Hz.  }
\label{fig:histoC} 
\end{figure}

In this appendix we show additional figures to complement the analysis presented in the main text. 
In Fig. \ref{fig:histoC} we show a histogram including the centroid frequencies of the 67 Hz QPOs reported in \cite{Belloni2013a}, save for the 2 peaks detected above 100 Hz. 

\begin{figure}
\includegraphics[width=0.43\textwidth]{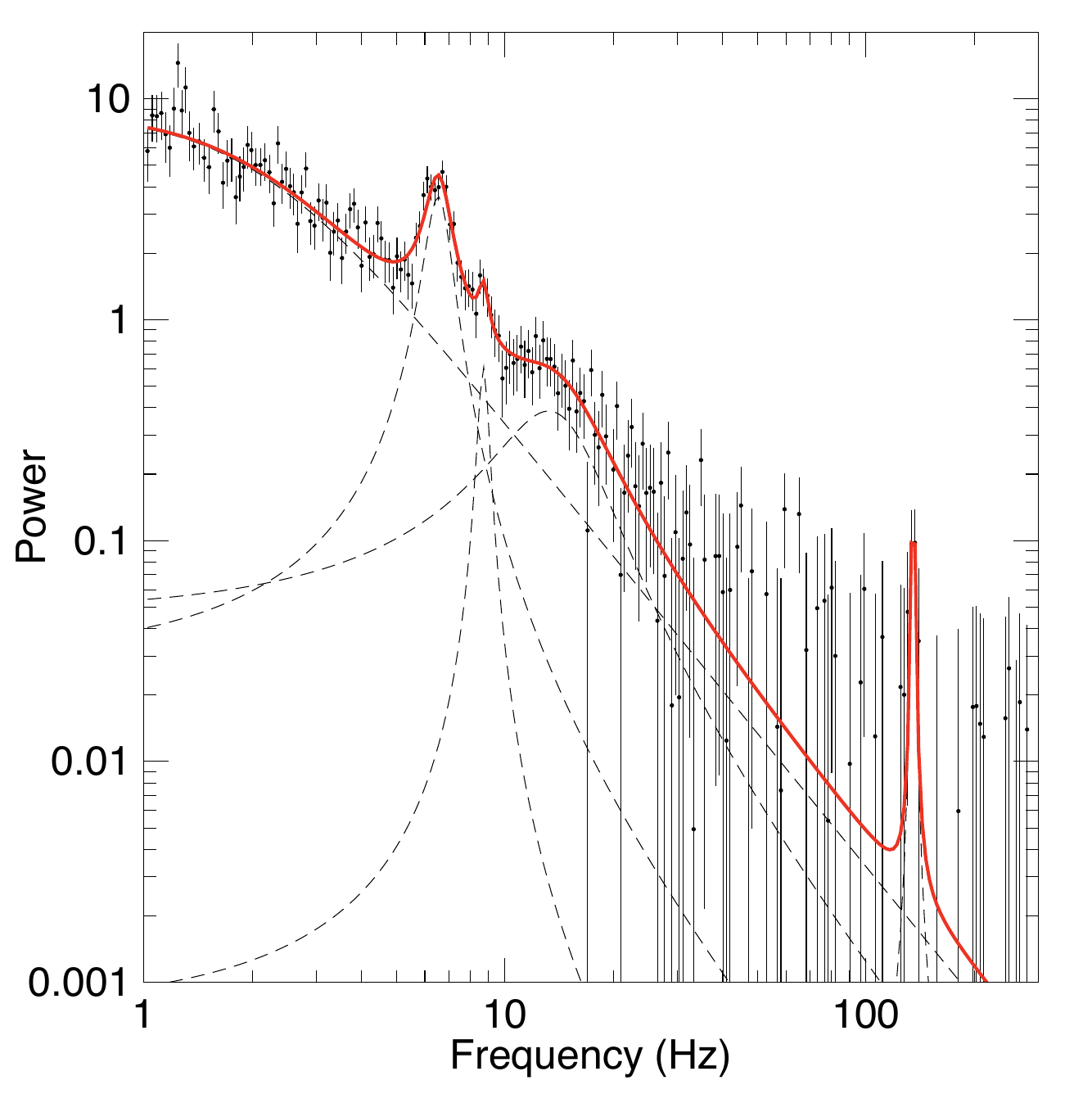}
\caption{A PDS from observation 50703-01-10-02, which includes a type-B QPO at 6.53$\pm$0.38 Hz, a type-C QPO at 8.71$\pm$0.09 Hz, and a high frequency QPO at 135.250$\pm$0.70.  }
\label{fig:bc}
\end{figure}

Figure \ref{fig:bc} shows the PDS from observation 50703-01-10-02, including a type-B, a type-C, and a HFQPO. The broad peak at the high-frequency side of the LFQPOs is centered at $\sim$13.2 Hz, and could be the hint of a second harmonic to the type-B QPO (albeit significantly broader than what would be expected in this case), or a L$_{h}$ component (see Fig. \ref{fig:PDS_components}). 

\begin{table*}
\caption{A summary of the quantities measured and derived in this work. Uncertainties are given at a 1-sigma level. The parameters marked by a $\dagger$ were fixed to the reported values. We note that the HFQPO has been reported in \cite{Belloni2013a} as a significant feature after taking into account the number of trials. }            
\label{tab:PDSpar}      
\centering                          
\begin{tabular}{c c c c c }        
\hline                        
Component                        &   Centroid Frequency           &      Width         &   Leahy normalisation   & rms [1-300 Hz]           \\
                         &   [Hz]                  &      [Hz]              &                  [counts/Hz]       & [1/Hz]                   \\     
\hline \hline
Broad component 1        &    0.0$^{\dagger}$      &    4.0$\pm$0.3         &    30$\pm$1           &    0.84$\pm$0.03                      \\     
Type-B QPO               &    6.53$\pm$0.05        &    1.2$\pm$0.1         &    6.4$\pm$0.5        &    0.19$\pm$0.01                      \\
Type-C QPO               &    8.7$\pm$0.1          &    0.6$^{\dagger}$     &    0.6$\pm$0.2        &    0.018$\pm$0.006                    \\
Upper HFQPO              &    135.1$^{+0.6}_{-0.8}$&    1.73112$^{+3}_{-1}$ &    0.7$\pm$0.2        &    0.20$\pm$0.006                     \\
Broad component 2        &    13.2$^{+0.7}_{-0.8}$ &   10$^{+3}_{-2}$       &    8.7$^{+0.9}_{-0.8}$&    0.157$\pm$$^{+0.027}_{-0.024}$     \\
\hline                    
\hline
$\chi^2/\nu$    &    \multicolumn{3}{c}{219.90/221} \\
\hline
\end{tabular}
\end{table*}

\end{appendix}

\end{document}